\documentclass[showpacs,prd,amssymb]{revtex4}
\usepackage{color}
\newcommand{\be}{\begin{equation}}
\newcommand{\ee}{\end{equation}}
\newcommand{\ba}{\begin{eqnarray}}
\newcommand{\ea}{\end{eqnarray}}

\begin{document}
\title{Mass Generation from Lie Algebra Extensions} \author{R. Aldrovandi}
\email{ra@ift.unesp.br} \author{V. C. de Andrade} \email{andrade@ift.unesp.br}
\author{A. L. Barbosa}
\email{analucia@ift.unesp.br} \author{J. G. Pereira}
\email{jpereira@ift.unesp.br} \affiliation{Instituto de F\'{\i}sica
Te\'orica\\
Universidade Estadual Paulista\\
Rua Pamplona 145 \\
01405-900 S\~ao Paulo SP \\ Brazil}

\begin{abstract}
Applied to the electroweak interactions, the theory of Lie algebra
extensions suggests a mechanism by which the boson masses are
generated without resource to spontaneous symmetry breaking.  It
starts from a gauge theory without any additional scalar field.  All
the couplings predicted by the Weinberg-Salam theory are present, and
a few others which are nevertheless consistent within the model.
\end{abstract}


\maketitle

\section{Introduction}

In the Weinberg--Salam model the boson masses are generated by spontaneous 
breakdown of the gauge symmetry, a process which hides invariance behind
the scene.  In pure gauge theories like QED and QCD, gauge
invariance is manifest in the forefront and there are no massive bosons.

The geometrical background of pure gauge theories is well known: a
principal bundle with spacetime for base space and the gauge group as
structure group~\cite{kn,aldp,nk}.  An essential feature of this
 structure is its direct product character: the bundle is
locally trivial.  This is to say that the fundamental vector fields
(which represent the group generators on the bundle)
 commute 
with the horizontal-lift vector fields (which represent a spacetime basis). Or still,
that the 1-form (the connection) that takes the fundamental fields back to the
corresponding generators in the group Lie algebra belongs to the adjoint representation
of the group.  Connections appear as interaction-mediating vector fields in Field
Theory: as the photon field (potential four-vector) in QED, as the gluon
fields in standard QCD. Such non-massive boson fields transform indeed like
connections, and  therefore preserve
the gauge invariance of the theory.

Electroweak theory, with its massive bosons, seems quite different
from QED and QCD from the geometrical point of view.  The massive
mediating fields do not transform like connections --- they actually
do not transform at all.  A natural question to be asked is: {\em What
is the geometrical background of electroweak theory?} An effort to
unveil the underlying geometry of electroweak interactions was
initiated years ago~\cite{Aldrovandi2}.  It was found that adding a
non--covariant piece to a connection led to field equations hinting
both at the possibility of mass generation and at effects of
gravitational character.  Some more steps in that way have been taken
recently~\cite{enlarged}, mainly in what concerns the gravitational
aspects.

The aim of this paper is to present an electroweak model whose
background geometry generates by itself the boson masses, with no
appeal to spontaneous breaking.  We intend, of course, to remain as
near as possible to the Weinberg--Salam (W-S) theory because of its
overwhelmingly successful phenomenology~\cite{cheng,greiner}.  Our model
also starts from a true gauge theory, but does without the additional
scalar field and its vacuum-modifying potential.  The basic point is that
the algebra of vector fields tangent to its principal bundle encapsulates
the whole local geometry of a gauge theory.  The new background referred
to (which we call an ``extended gauge model'') is obtained when that
algebra is conveniently modified by a procedure described by the theory
of Lie algebra extensions~\cite{Ald1}.  In the electroweak case, it is
applied to the Glashow algebra (GA), a Lie algebra obtained from the
generators of
$SU(2) \otimes U(1)$, but which incorporates the mixing angles in its
structure coefficients.  The resulting model has all the couplings
present in the W-S model, but also some which are absent.

The extended gauge formalism is presented in section \ref{2}.  That it
does lead to mass generation for the electroweak bosons is shown in
section
\ref{4}, but section \ref{3} gives a necessary, preparatory exposition
of the Glashow algebra ({\it GA}), which is the Lie algebra of the real
structure group of the model (Glashow group)~\cite{glashow}.  Some general
considerations on the model, including the problems still under
investigation, are made in the final section.

\section{ Extended Gauge Theories} \label{2}

\label{EGT} 
Let us begin by  presenting an extended gauge model.
In general terms, a connection
\begin{equation}
A_{\mu } = A^{a}{}_{\mu }X_{a} \label{connectiona}
\end{equation}
defines the covariant derivative
\begin{equation}
X_{\mu } = \partial _{\mu } - A_{\mu } , \label{change2}
\end{equation}
where $\{ X_{a} \}$ are the fundamental fields, and $\{ X_{\mu} \}$
the horizontal-lift fields \cite{Cho75a}.  A general basis on the fiber
bundle is defined by the set $\{X_{a},X_{\mu} \}$.  The geometric setup
is {\it locally} characterized by the corresponding commutation
relations,
\begin{eqnarray}
\nonumber \left[ X_{\mu },X_{\nu }\right] &=& -\;F^{a}{}_{\mu \nu
}X_{a},\\
\left[ X_{a},X_{\mu }\right] &=& 0 , \label{dirprod}\\
\nonumber \left[ X_{a},X_{b}\right] &=& f^{c}{}_{ab}X_{c}\ . 
\end{eqnarray}
The second commutator above declares the direct product character of
the bundle geometry and, furthermore, enforces the necessary adjoint
behavior of the connection $A$:
\begin{equation}
X_{a}(A^{b}{}_{\mu }) = f^{b}{}_{ca}A^{c}{}_{\mu }. \label{adjbehavior}
\end{equation}
Notice that we are working on the bundle manifold.  The usual
(non-homogeneous) derivative term in the transformation of $A^{a}{}_{\mu}$
only turns up when the expression above is pulled back to spacetime
\cite{Pop75}.  From the middle commutator in (\ref{dirprod}) and
(\ref{adjbehavior}) we obtain the expression for the field strength,
\begin{equation}
F^{a}{}_{\mu \nu } = \partial _{\mu }A^{a}{}_{\nu }-\partial _{\nu
}A^{a}{}_{\mu }+f^{a}{}_{bc}A^{b}{}_{\mu }A^{c}{}_{\nu }. \label{f}
\end{equation}
One of the Jacobi identities for the set of commutators gives
\begin{equation}
X_{a}(F^{b}{}_{\mu \nu })=f^{b}{}_{ca}F^{c}{}_{\mu \nu }.
\label{adjbehavior2}
\end{equation}
This condition shows that also $ F $ belongs to the adjoint
representation of the gauge group, whose generators are represented,
on the bundle, by the fields $X_{a}$.  In the language of Lie algebra
extensions, $F$ is called the ``non-linearity indicator''.  In the
direct product case, it coincides with the field strength of the gauge
field.

  The Jacobi identity for three fields $X_{\mu}$, $X_{\nu}$,
$X_{\rho}$ gives rise to the Bianchi identity. Gauge field dynamics can be
obtained by using the duality prescription: The sourceless field equations
are written just as the Bianchi identity, but applied to the dual of the
field strength.  The Yang--Mills equations come out:
\begin{equation}
X_{\mu }F ^{a\mu \nu }=0.  \label{gauge}
\end{equation}

An extended gauge theory comes forth when we break the direct product
in (\ref{dirprod}) through a change of basis \cite{enlarged},
\begin{equation}
X^{\prime }{}_{\mu } = X_{\mu }-B^{a}{}_{\mu }X_{a}, \label{Change3}
\end{equation}
which  is equivalent to
\begin{equation}
X^{\prime }{}_{\mu }=\partial_{\mu }-A^{\prime a}{}_{\mu }X_{a}
\label{dervgera}
\end{equation}
if we define
\begin{equation}
A^{\prime a}{}_{\mu } \equiv A^{a}{}_{\mu } + B^{a}{}_{\mu}.
\label{aab}
\end{equation}
The 1-form $A
^{\prime a}{}_{\mu }$ can be seen as a connection deformed by the
addition of a non--covariant form $B^{a}{}_{\mu}$ \cite{Aldrovandi2}.
Since the direct product is intimately related to the adjoint behavior
of a connection, it is necessary that neither $A^{\prime a}{}_{\mu }$
nor $B^{a}{}_{\mu}$ belong to the adjoint representation.  Expression
(\ref{dervgera}) leads to commutation relations of the form
\begin{eqnarray}
 \nonumber \left[ X^{\prime }{}_{\mu },X^{\prime }{}_{\nu }\right]
&=&-\;F^{\prime a}{}_{\mu \nu }X_{a},\\
\left[X_{a}, X^{\prime }{}_{\mu }\right] &=&C^{c}{}_{a \mu}X_{c},
\label{C''} \label{broken}\\ 
 \nonumber  \left[ X_{a},X_{b}\right] &=&f^{c}{}_{ab}X_{c}.\ 
\end{eqnarray}
The second commutator above gives the transformation law for the
object $ A^{\prime a}{}_{\mu }$ under the action of the gauge group:
\begin{equation}
X_{b}( A^{\prime a}{}_{\mu })=f^{a}{}_{cb} A^{\prime c}{}_{\mu
} - C^{a}{}_{b \mu}.
\label{misbehavior1}
\end{equation}
Comparison with (\ref{adjbehavior}) shows that $C^{a}{}_{b \mu}$
is the measure of its deviation from covariant behavior.   
We shall call the derivative (\ref{dervgera}), with the non-covariant
$A^{\prime a}{}_{\mu }$ in the position of a connection, a {\em
generalized derivative}.

The behavior of $B^{a}{}_{\mu}$ under the group action is obtained 
by using Eq. (\ref{Change3}) in the second commutator of Eqs.
(\ref{broken}):
\begin{equation}
X_{b}( B^{a}{}_{\mu })=f^{a}{}_{cb} B^{c}{}_{\mu
} - C^{a}{}_{b \mu}.
\label{misbehaviorb}
\end{equation}
The new non-linearity indicator $F^{\prime a}{}_{\mu \nu }$ can be
obtained from the first commutator:
\begin{equation}
F^{\prime a}{}_{\mu \nu }=\partial _{\mu }A^{\prime a}{}_{\nu }-\partial
_{\nu }A^{\prime a}{}_{\mu }+f^{a}{}_{bc}A^{\prime b}{}_{\mu }A^{\prime
c}{}_{\nu }-C^{a}{}_{c \mu }\ A^{\prime c}{}_{\nu }+\;C^{a}{}_{c \nu }\
A^{\prime c}{}_{\mu } \label{beta''} .
\end{equation}
Its behavior under the group action is fixed by the Jacobi
identity for two fields $X^{\prime }{}_{\mu }$ and one field $X_{a}$:
\begin{equation}
X_{b}F^{\prime a}{}_{\mu \nu }=f^{a}{}_{cb} F^{\prime c}{}_{\mu \nu }
- R^{\prime a}{}_{b\mu \nu },
\label{beta2}
\end{equation}
where
\begin{equation}
R^{\prime a}{}_{b\mu \nu } = X^{\prime }{}_{\mu }C^{a}{}_{b\nu }-X^{\prime
}{}_{\nu }C^{a}{}_{b\mu }-C^{a}{}_{d\mu }C^{d}{}_{b\nu }+C^{a}{}_{d\nu
}C^{d}{}_{b\mu }.  \label{curvtobe}
\end{equation}

Dynamics associated with algebra (\ref{C''}) is obtained by applying
the duality prescription to the Jacobi identity involving three fields
$X^{\prime }{}_{\mu }$.  The field equations turn out to be
\begin{equation}
X^{\prime }{}_{\mu }F^{\prime a\mu \nu }-C^{a}{}_{d\mu }F^{\prime d\mu \nu
}=0.  \label{eqmov2}
\end{equation}
These equation are, of course, linked to the choice of $C$, which is
constrained by a Jacobi identity for two fields $X_{a}$ and a field 
$X^{\prime}_{\mu}$:
\begin{equation}
X_{a}(C^{c}{}_{b \mu}) - X_{b}(C^{c}{}_{a \mu}) + f^{c}{}_{b
d}C^{d}{}_{a \mu} - f^{d}{}_{b a}C^{c}{}_{d \mu} - f^{c}{}_{a
d}C^{d}{}_{b \mu} = 0.
\label{Jacobi3}
\end{equation}

The term $R^{\prime a}{}_{b\mu \nu }$, which breaks the covariance of 
$F^{\prime a}{}_{\mu \nu }$, is constrained by an equation similar to 
the one satisfied by $C$:

\begin{equation}
X_{a}(R^{c}{}_{b \mu \nu}) - X_{b}(R^{c}{}_{a \mu \nu}) + f^{c}{}_{a
d}R^{d}{}_{b \mu \nu} + f^{d}{}_{b a}R^{c}{}_{d \mu \nu} - f^{c}{}_{b
d}R^{d}{}_{a \mu \nu} = 0.
\label{xr}
\end{equation}

We can infer by introducing (\ref{beta''})  and (\ref{beta2}) in
(\ref{eqmov2}) that a mass term for $ A^{\prime}$ can appear, a fact we
shall make profit of in section \ref{4}.  Thus, the basis change that
breaks the direct product can lead to a theory with massive vector fields
which, of course, behave no more like connections.  This is what happens
in the W-S model, here achieved through a different process.

The formalism just presented can be applied to any gauge group. 
Besides the same non-linearity indicator $F^{\prime a}{}_{\mu \nu }$
appearing in (\ref{C''}), an extra non--linear term $R^{\prime
c}{}_{a\mu \nu }$ turns up, whose aspect (\ref{curvtobe}) suggests a
curvature.  With some further elaboration it does lead to models of
gravitational type for 4-dimensional groups~\cite{enlarged}, but that
will not be our concern here. The counting of the degrees of freedom of
the theory will be  discussed at the end of section \ref{4}.


\section{Glashow Algebra and the Electroweak Interaction} \label{3}

In this section we present some basic concepts concerning the Glashow
Algebra~\cite{glashow}.  The main objective is to give a geometrical
role to the mixing angle in the electroweak theory.  The Glashow
Algebra GA is constructed as an extra support to the gauge theory, since
the introduction of the mixing angle emerges naturally in their
structure constants. In the usual W-S approach, the mixing angle is
introduced in order to diagonalize the mass matrix, the physical
fields appear as combinations of the original gauge potentials and
the underlying gauge algebra remains unchanged.

We start by considering a gauge theory and the direct product $SU(2)
\otimes U(1)$,
\begin{equation}
[ X_{a},X_{b}]=\epsilon^{c}{}_{ab}X_{c},
{\textnormal{ for }} a,b = 1, 2, 3
\label{algebrapd1}
\end{equation}
\begin{equation}
[ X_{a},X_{b}]=0,{\textnormal{ for }} a {\textnormal{ or }} b = 0.
\label{algebrapd2}
\end{equation}
The direct product $SU(2)\otimes U(1)$ leads to a sum of two gauge
theories, one abelian ($f^{a}{}_{b c} = 0$) and the other with gauge
potentials $A^a{}_{\mu}$ and field strength given by (\ref{f}), satisfying
respectively the transformations properties (\ref{adjbehavior}) and (\ref
{adjbehavior2}) with structure constants
$f^{a}{}_{b c} = \epsilon^{a}{}_{b c}$.  The Abelian and non-Abelian
sectors are quite independent. Furthermore,  there are no charged fields
in the algebraic schemes presented in section \ref{2}.  We know,
however, from the experimental data that there are two charged bosons
$W^{+}$ and $W^{-}$, and also that there is a mixture between the
Abelian and non--Abelian sectors giving an essential contribution to
the electron--positron cross section~\cite{mandl,greiner}. Thus, two
charged vector fields must be constructed and the algebra underlying
the theory must be modified to produce the necessary mixing.

Let us denote by $\{ \overline{X}_{a} \}$ a basis for the generators
of the new algebra, in terms of which the physical gauge potentials
will be written $\overline{A}_{\mu} = \overline{A}^{a}{}_{\mu}
\overline{X}_{a}$.  Since this time the change is only in the internal
sector, we must impose the following condition:
\begin{equation}
\overline{A}_{\mu} = \overline{A}^{a}{}_{\mu} \overline{X}_{a} 
= A^{a}{}_{\mu}{X}_{a} = A_{\mu}. 
\label{aabarra}
\end{equation}
That is, the spacetime sector remains unchanged.  Two charged gauge
fields and two neutral fields are constructed as a linear combinations
of the original ones:
\begin{equation}
\overline{A}^{1}{}_{\mu} = {\textstyle{\frac{1}{\sqrt{2}}}}
(A^{1}{}_{\mu} - iA^{2}{}_{\mu}),\label{an1}
\end{equation}%
\begin{equation}%
\overline{A}^{2}{}_{\mu} = {\textstyle{\frac{1}{\sqrt{2}}}}
(A^{1}{}_{\mu} + iA^{2}{}_{\mu}),\label{an2}
\end{equation}%
\begin{equation}
\overline{A}^{0}{}_{\mu} = \sin{\theta} A^{3}{}_{\mu} + \cos{\theta} 
A^{0}{}_{\mu},\label{an3}
\end{equation}
\begin{equation}
\overline{A}^{3}{}_{\mu} = \cos{\theta} A^{3}{}_{\mu} - \sin{\theta} 
A^{0}{}_{\mu},\label{an4}
\end{equation}
where $\theta$ is a mixing angle.
Using Eqs.  (\ref{aabarra}) and (\ref{an1})--(\ref{an4}) we obtain the
{\it GA} generators in terms of those of the direct product $SU(2)
\otimes U(1)$:
\begin{equation}
\overline{X}_{1} = {\textstyle{\frac{1}{\sqrt{2}}}} (X_{1} - 
iX_{2}),\label{x1}
\end{equation}
\begin{equation}
\overline{X}_{2} = {\textstyle{\frac{1}{\sqrt{2}}}} (X_{1} + 
iX_{2}),\label{x2}
\end{equation}
\begin{equation}
\overline{X}_{3} = \cos {\theta} X_{3} - \sin {\theta} 
X_{0},\label{x3}
\end{equation}
\begin{equation}
\overline{X}_{0} = \sin {\theta} X_{3} + \cos {\theta} 
X_{0}.\label{x0}
\end{equation}

The structure constants $\overline{f}^{c}{}_{ab}$ of the Glashow algebra 
\begin{equation}
[\overline{X}_{a},\overline{X}_{b}] = 
\overline{f}^{c}{}_{ab}\overline{X}_{c}
\end{equation}
are 
\[ 
\overline{f}^{0}{}_{12}= - i \sin \theta, \quad
\overline{f}^{3}{}_{12}= - i \cos \theta,
\]
\begin{equation}
\overline{f}^{1}{}_{10}=+i \sin \theta, \quad 
\overline{f}^{1}{}_{13}=+i \cos \theta,
\label{structure} 
\end{equation}
\[
\overline{f}^{2}{}_{23}= - i \cos \theta, \quad 
\overline{f}^{2}{}_{20}= - i \sin \theta.
\]
The mixture is in this way incorporated in the algebra through the
structure constants. The determinant of the related Killing-Cartan
bilinear form,
\begin{equation}
g_{ab} = \overline{f}^{c}{}_{ad} \overline{f}^{d}{}_{bc},
\end{equation}
is equal to zero, characterizing {\it GA} as a non-semisimple
algebra.  It can be shown that the physical fields (without mass), two
charged and two neutral, are indeed gauge fields, that is, they
transform like connections by the action of the group generators
$\overline{X}_{a}$.

The field strength associated to the physical fields is constructed
by appealing to the same arguments preceding Eq.  (\ref{aabarra}).  By
writing
\begin{equation}
\overline{F}^{a}{}_{\mu \nu }\overline{X}_{a} = F^{a}{}_{\mu \nu }X_{a},
\label{ffbarra}
\end{equation}
using Eqs.  (\ref{f}) and (\ref{aabarra}), and introducing the
coupling constant $g$, we arrive at
\begin{equation}
\overline{F}^{a}{}_{\mu \nu }=g\left[ \partial _{\mu }
\overline{A}^{a}{}_{\nu }- \partial_{\nu}\overline{A}^{a}{}_{\mu} +
g{}\overline{f}^{a}{}_{bc}\overline{A}^{b}{}_{\mu }
\overline{A}^{c}{}_{\nu}\right].
\label{fn}
\end{equation}
Expression (\ref{fn}) reflects a crucial result that we shall explore
from now on: All the expressions presented in section \ref{2} are
valid if we replace $A$ by $\overline{A}$, and the structure constants
$f$ by $\overline{f}$.  

The importance of {\it{GA}} is corroborated by the following example: 
It is possible to obtain from it the correct Lagrangian
for the massless electroweak theory.  This comes out by taking the
usual gauge Lagrangian
\begin{equation}
L = \frac{1}{8g^{2}}\int d^{3}x \, \textnormal{tr} \left( F_{\mu \nu
}F^{\mu \nu }\right) = \frac{1}{8g^{2}}\int d^{3}x 
\overline{F}^{a}{}_{\mu \nu} \overline{F}^{b \mu \nu} {\rm tr}
\left( \overline{X}_{a}^{*} \overline{X}_{b}^{*} \right),
\label{lag}
\end{equation}
with  a
representation $\{ \overline{X}_{a}^{*} \}$ whose non-vanishing traces~\cite{glashow} 
 are given by:
\begin{equation}
{\rm tr} \left( \overline{X}_{a}^{*}\overline{X}_{b}^{*}\right) =-2,\;\text{for }
(a,b)=(0,0),(1,2),(2,1)\;\text{e\ }(3,3).
\label{traco}
\end{equation}
The correct Lagrangian~\cite{mandl} appears after making the following
associations:
\begin{equation}
\begin{array}{c}
\overline{A}_{\hspace{0.1cm}\nu }^{1}\longrightarrow W_{\nu }^{-}, \\
\overline{A}_{\hspace{0.1cm}\nu }^{2}\longrightarrow W_{\nu }^{+}, \\
\overline{A}_{\hspace{0.1cm}\nu }^{3}\longrightarrow Z_{\nu }, \\
\overline{A}_{\hspace{0.1cm}\nu }^{0}\longrightarrow A_{\nu }.
\end{array}\label{array1}
\end{equation}
 
 Though we have presented the Lagrangian for the massless electroweak
 theory, we proceed using the formalism of equations of motion since
 the theory of extended Lie algebras works directly with that
 formalism, as can be seen from Eq.  (\ref{eqmov2}).

\section{Generation of Mass} {\label{4}}

\subsection{Equation of Motion}

As seen in section \ref{2} an extended gauge theory can be obtained 
by adding to a connection a non-Covariant part:
\[
\overline{A}^{\prime c}{}_{\mu} = \overline{A}^{a}{}_{\mu}
+\overline{B}^{a}{}_{\mu }.
\]
$\overline{A}^{\prime c}{}_{\mu }$ will
be interpreted as the physical massive fields, that is, they will be
identified as
\begin{equation}
\begin{array}{c}
\overline{A}^{\prime 1}{}_{\nu }\longrightarrow W_{\nu }^{-}, \\
\overline{A}^{\prime 2}{}_{\nu }\longrightarrow W_{\nu }^{+}, \\
\overline{A}^{\prime 3}{}_{\nu }\longrightarrow Z_{\nu }, \\
\overline{A}^{\prime 4}{}_{\nu }\longrightarrow A_{\nu}.
\end{array}
\end{equation}
$\overline{B}^{a}{}_{\mu }$ will be responsible for
the mass generation and, if so wished, for the introduction of a scalar
field, a candidate Higgs field.

The mass terms are obtained by exploring the  extended gauge
dynamics of {\it{GA}}.  
The equation of motion is now
\begin{equation}
\overline{X}^{\prime}_{\mu }\overline{F}^{\prime a\mu \nu } -
\overline{C}^{a}{}_{d\mu } \overline{F}^{\prime d\mu \nu } = 0,
\label{eqmov1}
\end{equation}
with $\overline{X}^{\prime}_{\mu }=\partial_{\mu}-\overline{A}^{\prime 
a}{}_{\mu}\overline{X}_{a}$ and $\overline{X}_{a}\overline{A}^{\prime 
b}{}_{\mu}=\overline{f}^{b}{}_{ca}\overline{A}^{\prime 
c}{}_{\mu}-\overline{C}^{b}{}_{a \mu}$.
This  expression can be rewritten with the help of (\ref{beta2}) as
\[
\partial {}_{\mu }\overline{F}^{\prime a\mu \nu }-\overline{f}^{a}{}_{cb}
\overline{F}^{\prime c\mu \nu }\overline{A}^{\prime b}{}_{\mu }{} +
\overline{A}^{\prime b}{}_{\mu }\overline{R}^{\prime a}{}_{b}{}^{\mu \nu
}-\overline{C}^{a}{}_{d\mu } \overline{F}^{\prime d\mu \nu }=0,
\]
with
\begin{equation}
\overline{F}^{\prime a}{}_{\mu \nu } = \partial_{\mu
}\overline{A}^{\prime a}{}_{\nu} - \partial _{\nu
}\overline{A}^{\prime a}{}_{\mu } + \overline{f}^{a}{}_{bc}
\overline{A}^{\prime b}{}_{\mu} \overline{A}^{\prime c}{}_{\nu} -
\overline{C}^{a}{}_{c\mu }\ \overline{A}^{\prime c}{}_{\nu } +
\overline{C}^{a}{}_{c\nu }\overline{A}^{\prime c}{}_{\mu },
\end{equation}
and
\begin{equation}
\overline{R}^{\prime a}{}_{b}{}^{\mu \nu } = \overline{X}^{\prime \mu} 
\overline{C}^{a}{}_{b}{}^{\nu}-\overline{X}^{\prime \nu} 
\overline{C}^{a}{}_{b}{}^{\mu}-\overline{C}^{a}{}_{e}{}^{\mu 
}\overline{C}^{e}{}_{b}{}^{\nu}+\overline{C}^{a}{}_{e}{}^{\nu
}\overline{C}^{e}{}_{b}{}^{\mu}.  \label{curvtobe2}
\end{equation}

We shall henceforth simplify the notation to agree with that of section 
\ref{2}, dropping the  bars from fields  and structure
constants. In terms of the broken potencials ${A}^{\prime a}{}_{\mu }$,
Eq. (\ref{eqmov1}) can be rewritten as
\[
\partial ^{\mu }\partial _{\mu }A^{\prime a \nu }-\partial _{\mu
}\partial^{\nu }A^{\prime a \mu }+\left[ f^{a}{}_{bc}A^{\prime c \nu
}+C^{a}{}_{b}{}^{\nu }\right] \partial _{\mu }A^{\prime b \mu } +
\left[ 2f^{a}{}_{bc}A^{\prime b}{}_{\mu }-2C^{a}{}_{c}{}^{\mu }\right]
\partial _ {\mu}A^{\prime c \nu }
\]
\begin{equation}
+\left[ \partial _{\mu }C^{a}{}_{c}{}^{\nu }-C^{a}{}_{d\mu
}C^{d}{}_{c}{}^{\nu }\right] A^{\prime c \mu }+ A^{\prime b}{}_{\mu
}R^{\prime a}{}_{b}{}^{\mu \nu }+\left[ C^{a}{}_{c\mu
}+f^{a}{}_{cb}A^{\prime b}{}_{\mu }\right] \partial^{\nu}A^{\prime c
\mu }\label{eqmov3}
\end{equation}
\[
+\left[f^{a}{}_{cb}C^{c}{}_{d\mu }-f^{c}{}_{bd}C^{a}{}_{c\mu }\right]
A^{\prime b \mu}A^{\prime d \nu} - f^{a}{}_{cb}C^{c}{}_{d}{}^{\nu
}A^{\prime b}{}_{\mu}A^{\prime d \mu }
-f^{a}{}_{cb}f^{c}{}_{de}A^{\prime b}{}_{\mu }A^{\prime d \mu
}A^{\prime e \nu }
\]
\[
+ \left[-\partial _{\mu }C^{a}{}_{d}{}^{\mu} + C^{a}{}_{c\mu
}C^{c}{}_{d}{}^{\mu }\right] A^{\prime d \nu } = 0.
\]
For $C=0$ this gives just the Yang-Mills equations.  Let us therefore
examine the non trivial part, with the terms containing $C's$:
\[
C^{a}{}_{b}{}^{\nu }\partial _{\mu }A^{\prime b \mu } - 
2C^{a}{}_{c\mu }\partial_{\mu }A^{\prime c\nu} + 
C^{a}{}_{c\mu }\partial^{\nu}A^{\prime c \mu}
\]
\begin{equation}
+ \left[ \partial _{\mu }C^{a}{}_{c}{}^{\nu }-C^{a}{}_{d\mu
}C^{d}{}_{c}{}^{\nu }+R^{\prime a}{}_{c\mu}{}^{\nu }\right] A^{\prime
c\mu } +\left[ f^{a}{}_{cb}C^{c}{}_{d\mu }-f^{c}{}_{bd}C^{a}{}_{c\mu
}\right] A^{\prime b\mu }A^{\prime d\nu }
\label{cpart}
\end{equation}
\[
- f^{a}{}_{cb}C^{c}{}_{d}{}^{\nu}A^{\prime b}{}_{\mu }A^{\prime d \mu}
+ \left[-\partial _{\mu }C^{a}{}_{d}{}^{\mu} + C^{a}{}_{c\mu}
C^{c}{}_{d}{}^{\mu }\right] A^{\prime d \nu}.
\]
We see that the last two terms, which we indicate
\begin{equation}
{{\bf Q}^{a\nu}} = \left[ -\partial _{\mu }C^{a}{}_{d}{}^{\mu
}+C^{a}{}_{c\mu }C^{c}{}_{d}{}^{\mu }\right] A^{\prime d \nu },
\label{Q}
\end{equation}
can provide a mass term for the component $a$ of $A^{\prime}$,
given by the term with $d = a$ of the sum over $d$.  In fact ${{\bf
Q}^{a\nu }}$, besides being responsible for the masses, will give
rise to coupling terms.  As our main purpose is to obtain the
phenomenologically correct values for the masses, we begin by
analyzing (\ref{Q}) for each component of $A^{\prime d \nu }$.

An initial point to consider is the experimental fact that one of the
neutral bosons remains massless, that is, it behaves like a
connection.  This means that the $C$ related to it must be zero.  In fact,
up to this point, there is no difference between $A^{\prime
0}{}_{\nu}$ (or $A^{\nu }$) and $A^{\prime 3}{}_{\nu}$ (or $Z^{\nu
}$).  We shall make the choice $C^{0}{}_{a\mu }\equiv 0$, which will
break the symmetry between them.  Thus, $A^{\prime 0 \nu} = A^{\nu}$
will be the photon field.  Its transformation under the Glashow group
is
\begin{equation}
X_{a}A^{\prime 0 \nu }=f^{0}{}_{ca}A^{\prime c \nu },
\end{equation}
which means
\begin{equation}
C^{0}{}_{c \mu } \equiv 0, \;\; \forall \, c \;\; {\textnormal{and}}
\;\; \mu.
\end{equation}
It also implies $B^{0}{}_{\mu }\equiv 0$.

Electric charge conservation imposes additional conditions.  A term by
term examination of Eq.  (\ref{cpart}) for each component shows
that many components of $C$ must be made to vanish.  The only
nonvanishing components are four: $C^{1}{}_{1\mu }$, $C^{2}{}_{2\mu
}$, $C^{3}{}_{0\mu }$ and $ C^{3}{}_{3\mu}$.  Up to this point these
$C's$ are arbitrary but, as we are going to see, they will have to
assume some special forms in order to generate the correct mass
values.

Let us proceed to analyze ${{\bf Q}^{a\nu }}$, 
keeping only the four components of $C$ above. Writing
\begin{equation}
{{\bf Q}^{a\nu }}=Q^{a}{}_{d}A^{\prime d \nu}
\end{equation}
with
\begin{equation}
Q^{a}{}_{d}=\left[ -\partial _{\mu }C^{a}{}_{d}{}^{\mu }+C^{a}{}_{c\mu
}C^{c}{}_{d}{}^{\mu }\right],
\label{q}
\end{equation}
we obtain:
\[
\begin{array}{cc}
{\bf{Boson}} \; \; \; \; \;  & {\bf{Q^{a \nu}}} \\ 
W^{-}_{\mu} \; \; \; \; \; & {\bf{Q^{1 \nu}}}=(-
\partial_{\mu}C^{1}{}_{1}{}^{\mu} + C^{1}{}_{1
\mu}C^{1}{}_{1}{}^{\mu})W^{- \nu} \\
W^{+}_{\mu} \; \; \; \; \; & {\bf{Q^{2 \nu}}}=(-
\partial_{\mu}C^{2}{}_{2}{}^{\mu} + C^{2}{}_{2
\mu}C^{2}{}_{2}{}^{\mu})W^{+ \nu} \\
Z_{\mu} \; \; \; \; \; & {\bf{Q^{3 \nu}}}=(-
\partial_{\mu}C^{3}{}_{3}{}^{\mu} + C^{3}{}_{3
\mu}C^{3}{}_{3}{}^{\mu})Z^{\nu} -
\partial_{\mu}C^{3}{}_{0}{}_{\mu}A^{\nu} \\ A_{\mu} \; \; \; \; \; &
{\bf{Q^{0 \nu}}}=0
\end{array}
\]

The masses $m_{W}$ and $m_{Z}$ of 
$W^{-}$($W^{+}$) and $Z$ are known.  To match their values we must choose
some model for the $C$'s.

\subsection{Restricting the Model}

Our model aims to recover all the predictions of W-S theory without
remaining restricted to them.  This means that we leave it open to the
possibility of new couplings.  It is crucial to remember that $C$
accounts also for the non-covariance of $B$,
\begin{equation}
C^{a}{}_{b\mu }=-X_{b}(B^{a}{}_{\mu })+f^{a}{}_{cb}B^{c}{}_{\mu }
\label{c}
\end{equation}
with $B$ arbitrary.  We adopt for
$B^{c}{}_{\mu }$ an expression as general as possible at this point,
also contemplating simplicity.  It must contain one term independent
of the coordinates $x^{\mu}$ to originate masses, and a linear part
that will be related to the candidate Higgs field.  We write then
\begin{equation}
B^{a}{}_{\mu }=\alpha M^{a}{}_{\mu }+\beta K^{a}{}_{\mu }(x^{\mu }),
\label{modelob}
\end{equation}
with $\alpha$, $\beta$ real numbers.  Substituting (\ref{modelob}) in
(\ref{c}) and considering the structure constants (\ref{structure}) we
get
\[
C^{1}{}_{1\mu } = -\alpha X_{1}(M^{1}{}_{\mu })-\beta X_{1}(K^{1}{}_{\mu
})-i\cos \theta \left[ \alpha M^{3}{}_{\mu }+\beta K^{3}{}_{\mu 
}\right],
\]
\[
C^{2}{}_{2\mu } = -\alpha X_{2}(M^{2}{}_{\mu })-\beta
X_{2}(K^{2}{}_{\mu })+i\cos \theta \left[ \alpha M^{3}{}_{\mu }+\beta
K^{3}{}_{\mu }\right],
\]
\begin{equation}
C^{3}{}_{3\mu } = -\alpha X_{3}(M^{3}{}_{\mu })-\beta
X_{3}(K^{3}{}_{\mu }), \label{tab1}
\end{equation}
\[
C^{3}{}_{0\mu }  = -\alpha X_{0}(M^{3}{}_{\mu })-\beta X_{0}(K^{3}{}_{\mu
})
\]
and we can evaluate ${\bf{Q}}^{a \nu}$ for each particle.

\subsubsection{Photon}

Since we have in this case a connection, $B^{0}{}_{\mu }\equiv 0$ and 
$C^{0}{}_{a\mu }\equiv 0$. In consequence,
\begin{equation}
{{\bf Q}^{a\nu }}_{photon} = 0,
\end{equation}
and there is no mass generation for the photon.

\subsubsection{Z}

For the third component, we have 
\begin{equation}
{{\bf Q}^{3\nu }}_{Z}=Q^{3}{}_{3}Z^{\nu }+ Q^{3}{}_{0}A^{\nu }
\end{equation}
with
\[
Q^{3}{}_{3} = \alpha ^{2}X_{3}(M^{3}{}_{\mu })X_{3}(M^{3\mu })+2\alpha
\beta X_{3}(M^{3}{}_{\mu })X_{3}(K^{3\mu}) +
\]
\begin{equation}
\beta^{2}X_{3}(K^{3}{}_{\mu
})X_{3}(K^{3\mu })+\beta X_{3}(\partial _{\mu }K^{3\mu}),
\label{qz} 
\end{equation}
and
\begin{equation}
Q^{3}{}_{0} = \beta X_{0}(\partial _{\mu }K^{3\mu}).
\label{qzo} 
\end{equation}
The first term in (\ref{qz}) must be the mass term up to a sign.  We
must have then
\begin{equation}
\alpha ^{2}X_{3}(M^{3}{}_{\mu })X_{3}(M^{3\mu })=-m_{Z}^{2}.\label{alpha2}
\end{equation}
One possible solution comes from the condition
\begin{equation}
X_{3}(M^{3}{}_{\mu })= \pm \frac{i}{2 \alpha }m_{Z}I_{\mu }, \label{v1}
\end{equation}
where  $I_{\mu }$ is a row-vector satisfying $I_{\mu }I^{\mu } = 4$.

The term quadratic in $\beta$ of Eq. (\ref{qz}) corresponds to an interaction of the 
field $Z$ with a field $\sigma(x)$ up to a constant $D$:
\begin{equation}
\beta ^{2}X_{3}(K^{3}{}_{\mu })X_{3}(K^{3\mu}) = D^{2}\sigma^{2},
\label{zs}
\end{equation}
which leads to the condition
\begin{equation}
X_{3}(K^{3\mu}) = \pm \frac{D \sigma }{\beta} \, I^{\mu}.
\label{kd}
\end{equation}

Finally, using conditions (\ref{v1}) and (\ref{kd}), the second and 
fourth terms in (\ref{qz}) are obtained. The second term is
\begin{equation}
2\alpha \beta X_{3}(M^{3}{}_{\mu })X_{3}(K^{3\mu })= 2 i D \sigma m_{Z}.
\label{ab0}
\end{equation}
It is remarkable that the
two terms (\ref{alpha2}) and (\ref{zs}) imply, in our formalism, the
presence of (\ref{ab0}), a type of coupling which is also present in the W-S model.
The fourth term corresponds to
\begin{equation}
\beta X_{3}(\partial _{\mu }K^{3\mu }) = \pm D \left(
\partial _{\mu }\sigma \right) I^{\mu },
\label{nt}
\end{equation}
which shows that our model contains a new coupling, absent in
the W-S model: A derivative $\sigma$ field term coupled with $Z^{\nu}$.

It is possible to choose $D$ so that we have the same couplings of 
W-S with their constants. Let us begin by matching the linear coupling 
in Eq. (\ref{ab0})
\begin{equation}
2 i D \sigma m_{Z}= -\frac{g\sigma m_{Z}}{
\cos \theta_{W}}
\label{ab1}
\end{equation}
we then have 
\begin{equation}
D = \frac{ig\sigma }{2\beta \cos \theta_{W}}
\label{d}
\end{equation}
where $g = e\, \sin{\theta_{W}}$ is a coupling constant, $\theta_{W}$ is the 
Weinberg angle and $\sigma$ represents the Higgs field.

Taking Eq. (\ref{d}) into (\ref{zs}), we obtain exactly the
$Z-\sigma^{2}$ interaction term of the W-S model, 
\begin{equation}
\beta ^{2}X_{3}(K^{3}{}_{\mu })X_{3}(K^{3\mu}) = -\frac{g^{2}\sigma^{2}}{
4\cos \theta_{W}},
\label{v2}
\end{equation}

The term (\ref{nt}) becomes:
\begin{equation}
\beta X_{3}(\partial _{\mu }K^{3\mu })=\frac{ig}{2\cos \theta_{W}}\left(
\partial _{\mu }\sigma \right) I^{\mu },
\end{equation}
the above mentioned non-standard term in our model.
Here, the question arises whether the mixing angle introduced by the 
structure constants of GA coincides with the Weinberg 
angle. The answer is positive. The equality is necessary if we 
want to match the coupling terms.

Summing up our results, we have obtained in the field equation the
following terms:
\begin{equation}
{{\bf Q}^{3 \nu}}_{Z}=\left[ -m_{Z}^{2}-\frac{g^{2}\sigma ^{2}}{4\cos 
\theta_{W}}--\frac{g\sigma m_{Z}}{
\cos \theta_{W}}+\frac{ig}{2\cos \theta_{W}}\left(
\partial _{\mu }\sigma \right) I^{\mu }\right] Z^{\nu } + \beta 
X_{0}(\partial_{\mu }K^{3\mu })A^{\nu}.
\label{qz3}
\end{equation}
This coincides with the W-S model except for the presence of the last
two terms.  These are theoretically consistent within the model, but
their eventual measurable effects are still to be evaluated.

\subsubsection{W$^{-}$ and W$^{+}$}
Once we have learned that it is possible to recover the W-S model
by choosing correctly the free parameters of our model, we make from now
on direct contact with that model. As in the previous case, we write 
\begin{equation}
{{\bf Q}^{1\nu}}_{W^{-}}=Q^{1}{}_{1} \ W^{-\nu }
\end{equation}
with
\[
Q^{1}{}_{1} =\alpha^{2} 
\left[ X_{1} ( M^{1}{}_{\mu}) + i \cos{\theta_{W}}M^{3}{}_{\mu} \right]^{2} +
\]
\[
2\alpha \beta \left\{ X_{1}(M^{1}{}_{\mu })X_{1}(K^{1\mu 
})+i\cos{\theta_{W}}\left[
K^{3}{}_{\mu }X_{1}(M^{1\mu })
+ M^{3}{}_{\mu
}X_{1}(K^{1\mu }\right] -\cos ^{2}{\theta_{W}} M^{3}{}_{\mu }K^{3\mu } 
\right\} + 
\]
\[
\beta^{2} \left[ X_{1}(K^{1}{}_{\mu})+i\cos{\theta_{W}} K^{3}{}_{\mu } 
\right]^2 
+\beta \left[ X_{1}(\partial _{\mu }K^{1 \mu})+i\cos{\theta_{W}} 
\partial _{\mu}K^{3\mu} \right].
\]
For the mass term we have
\[
- \alpha ^{2}\left[ \cos \theta_{W} M^{3}{}_{\mu }-iX_{1}(M^{1}{}_{\mu })
\right] ^{2}=-m_{W}^{2},
\]
corresponding to
\begin{equation}
X_{1}(M^{1}{}_{\mu })=\pm \left( \frac{im_{W}}{2\alpha }I_{\mu 
}\right) -i\cos \theta_{W}
M^{3}{}_{\mu }.  \label{v3}
\end{equation}
The term quadratic in $\beta$ matched to W-S coupling leads to 
\[
- \beta ^{2}\left[ \cos \theta_{W} K^{3}{}_{\mu }-iX_{1}(K^{1}{}_{\mu })
\right] ^{2}=-\frac{1}{4}g^{2}\sigma ^{2},
\]
and we have the condition
\begin{equation}
X_{1}(K^{1}{}_{\mu })= \pm \left( \frac{ig\sigma }{4\beta }I_{\mu }\right) -i\cos \theta_{W}
K^{3}{}_{\mu }.  \label{v4}
\end{equation}

It is important to notice that the 
signs in Eqs. (\ref{v3}) and (\ref{v4}) are defined 
independently from each other. Thus we may fit them to obtain from 
the term
proportional to $\alpha \beta $ the linear
term in the Higgs field present in the W-S model:
\[
2\alpha \beta \lbrack X_{1}(M^{1}{}_{\mu })X_{1}(K^{1\mu })+i\cos \theta_{W}
K^{3}{}_{\mu }X_{1}(M^{1\mu })
\]
\[
+i\cos \theta_{W} M^{3}{}_{\mu }X_{1}(K^{1\mu })-\cos ^{2}\theta_{W}
M^{3}{}_{\mu }K^{3\mu }]=-gm_{W}\sigma
\]
with 
\begin{equation}
X_{1}(M^{1}{}_{\mu }) = \frac{im_{W}}{2\alpha }I_{\mu }-i\cos \theta_{W}
M^{3}{}_{\mu }
\end{equation}
and
\begin{equation}
X_{1}(K^{1}{}_{\mu }) = \frac{ig\sigma }{4\beta }I_{\mu }-i\cos \theta_{W}
K^{3}{}_{\mu } .
\end{equation}
As in the previous case, we get also an extra term:
\[
\beta \left[ X_{1}(\partial _{\mu }K^{1\mu })+i\cos \theta_{W} \left(
\partial _{\mu }K^{3\mu }\right) \right] =\frac{ig}{2}\left( \partial
_{\mu }\sigma \right) I^{\mu }.
\]
Gathering terms we obtain
\begin{equation}
{{\bf Q}^{1\nu}}_{W^{-}}=\left[ -
m_{W}^{2}-\frac{1}{4}g^{2}\sigma^{2}-gm_{W}\sigma+\frac{ig}{2}\left(
\partial _{\mu }\sigma \right) I^{\mu }\right]W^{-\nu},
\label{termw}
\end{equation}
possessing the mass term, the expected coupling with the Higgs field,
and a new, non-standard derivative coupling.

In the same way, evaluating ${{\bf Q}^{2\nu}}_{W^{+}}$, we obtain the
following conditions:
\begin{equation}
X_{2}(M^{2}{}_{\mu })= - \frac{im_{W}}{2\alpha }I_{\mu }+i\cos \theta_{W}
M^{3}{}_{\mu } 
\label{xm2}
\end{equation}
and
\begin{equation}
X_{2}(K^{2}{}_{\mu }) = - \frac{ig\sigma }{4\beta }I_{\mu }+i\cos \theta_{W}
K^{3}{}_{\mu},\label{xk2}
\end{equation}
leading to
\begin{equation}
{{\bf Q}^{2\nu}}_{W^{+}}=\left[ -m_{W}^{2}-\frac{1}{4}g^{2}\sigma
^{2}-gm_{W}\sigma-\frac{ ig}{2}\left( \partial _{\mu
}\sigma \right) I^{\mu }\right] W^{+ \nu}.
\end{equation}

The signs of Eqs. (\ref{xm2}) and (\ref{xk2}) are chosen, on one hand, to
fit those of W-S model; on the other hand, to be consistent with
the equality of the $W^{+}$ and $W^{-}$ masses. This implies, from 
the definition of ${\bf{Q^{2 \nu}}}$ in the table below Eq. (\ref{q}) 
and from the expressions of $C^{1}{}_{1\mu }$ and $C^{2}{}_{2\mu }$
 in (\ref{tab1}), that we must 
have
\begin{equation}
C^{1}{}_{1\mu } = -\ C^{2}{}_{2\mu } .
\label{c1c2}
\end{equation}
A positive sign in the left-hand side 
would lead to $C^{3}{}_{3\mu } = 0$. 

For completeness, we exhibit the components of $C$ determined by the
model:
\begin{eqnarray*}
C^{1}{}_{1\mu } &=&- C^{2}{}_{2\mu }=-\frac{i}{2}\left[
m_{W}+\frac{g\sigma }{2}\right] I_{\mu }, \\ C^{3}{}_{3\mu }
&=&-\frac{i}{2}\left[ m_{Z}+\frac{g\sigma }{2\cos \theta_{W} }\right]
I_{\mu }.
\end{eqnarray*}
Notice that, using the relation $m_{Z}=m_{W}/\cos
\theta_{W}$, we find $C^{3}{}_{3\mu }={C^{1}{}_{1\mu }}/{\cos
\theta_{W}}$.  The component $C^{3}{}_{0\mu }$ is up
to now completely arbitrary.
Loosely speaking, $C$, the object which measures the covariance breaking of 
$A^{\prime}_{\mu}$, is directly related to mass generation and to the 
existence of another field which we are associating to the Higgs field. 

A balance of the degrees of freedom should be done. Firstly, we notice 
that in the very beginning of the process of adding a non-covariant 
part to the connection, we have 3 degrees of freedom for $B^{a}{}_{\mu}$. They come
from the three non-null gauge component, each one with two degrees of freedom (it is a
massless vector term) from which we subtract three degrees of freedom due to the
constraints (\ref{misbehaviorb}) on $B^{a}{}_{\mu}$. Now, adding the eight degrees of
freedom for the massless fields $A^{a}{}_{\mu}$ of the theory, it totalizes eleven.
The same total number of degrees of freedom is computed after the process of mass
generation, since we have three massive $A^{\prime c}{}_{\mu}$, amounting to
nine degrees of freedom, plus the boson that remains massless, with 2
degrees of freedom. Notice finally that the degree of freedom corresponding to 
the candidate Higgs field $\sigma$ is already included in those of B's.
   
\section{Conclusions and Final Comments}
{\label{6}}

We have presented a procedure to generate masses for the bosons in
electroweak theory which is alternative to spontaneous symmetry
breaking.  The method takes its roots in the theory of Lie algebra
extensions, applied in the case to the Glashow algebra.  The extension
of a Lie algebra is another Lie algebra, so that new Jacobi identities
appear.  One of them leads to a new Bianchi identity.  The dynamic
equations for the boson fields are obtained by applying the duality
prescripition to that Bianchi identity.  The formalism leads, in this
way, directly to the field equations.  It should be recalled that
quantization, despite the modern heavy reliance on Lagrangians and
some statements to the contrary, can be realized directly from the
field equations \cite{KYFAK,bjorken}.

Working only with the equations of motion, we have shown that
it is possible to obtain the correct masses for $W^{+}$, $W^{-}$ and $Z$,
while keeping a fourth boson $A$ massless.  The model predicts  all
the bosonic couplings present in W--S model. 

Another feature of our model is the introduction of a 
scalar field $\sigma$, candidate (so called because its dynamics is still
under examination) to play the role of the Higgs field of the W--S model.
Besides the  Higgs-boson  couplings of the W--S model, 
four non-standard couplings turn up.  The latter
are consistent within the model and their contributions to cross sections
are under study. The presence of
$\sigma$ field is necessary to have the same number the degrees of
freedom
 before and after mass generation. It is also directly linked to the
coefficients
$C^{a}{}_{b \mu}$,  which measure the direct
product breaking responsible for the appearence of the masses.

The $\sigma$ field dynamics,
the renormalizability of the extra couplings as well as the Lagrangian
formalism, are still under study. The same is true of the
gravitational counterpart of the model~\cite{avrz2}.
 
\begin{acknowledgments}
The authors thank FAPESP (Brazil) and CNPq (Brazil) for financial
support.  They also thank B. M. Pimentel for very fruitful
discussions.
\end{acknowledgments}



\begin{thebibliography}{30}

\bibitem{kn}
S. Kobayashi and K. Nomizu, {\it Foundations of
Differential Geometry, 1st vol.} (Interscience, New York, l963).

\bibitem{aldp}
R. Aldrovandi and J.G. Pereira, {\it An Introduction
to Geometrical Physics} (World Scientific, Singapore, 1995).

\bibitem{nk}
M. Nakahara, {\it Geometry, Topology and Physics} ( IOP Publishing,  
Bristol, 1990).

\bibitem{Aldrovandi2}
R. Aldrovandi, Phys.  Lett. {\bf 
A155}, 459 (1991).
    
\bibitem{enlarged}
R. Aldrovandi  and A.L. Barbosa, Int. J. Theor.  Phys. {\bf
39}, 2779 (2000).

\bibitem{cheng} T.P. Cheng and L.F. Li, {\it Gauge Theory of
Elementary Particles} (Oxford University Press, Oxford, 1984).

\bibitem{greiner}
W. Greiner and B. M\"uller,  {\it{Gauge Theory of Weak
Interactions}} (Springer, New York, 1996).

\bibitem{Ald1}
R. Aldrovandi, J. Math.  Phys. {\bf 32}, 2503 (1991).

\bibitem{glashow}
A.L. Barbosa, Int.  J. Theor.  Phys. {\bf 39}, 1985 (2000).

\bibitem{Cho75a}
Y.M. Cho, Phys. Rev.  {\bf D12}, 3789 (1975).

\bibitem{Pop75}
D.A. Popov, Theor. Math. Phys.  {\bf 24}, 347 (1975).

\bibitem{mandl}
F. Mandl and G. Shaw, {\it{Quantum Field Theory}} (John Wiley,
New York, 1984).



\bibitem{KYFAK}
G. K\"allen, Ark.  Fys {\bf 2} (1950) 187 and 371; C.N. Yang and D.
Feldman, Phys.  Rev.  {\bf 79}, (1950) 972; R. Aldrovandi and R.
Kraenkel, J. Math.  Phys.  {\bf 30}, (1989) 1866.

\bibitem{bjorken} 
J.D. Bjorken and S.D. Drell, {\it{ 
Relativistic Quantum Mechanics}} (McGraw--Hill, New York, 1964).

\bibitem{avrz2}
R. Aldrovandi, V.C. Andrade, A.L. Barbosa, and J.G. Pereira,
{\it{Gravitational Model from Extended Gauge Theories}} (in
preparation).

\end{thebibliography}
\end{document}